**Taxonomy and Jargon in SETI as an Interdisciplinary Field of Study.** J. T. Wright,[1] [1]Department of Astronomy and Astrophysics and Center for Exoplanets and Habitable Worlds, 525 Davey Laboratory, The Pennsylvania State University, University Park, PA 16802

**Overall Goal(s) and Objective(s):**

While SETI is often thought of as a part of radio astronomy with optical SETI, artifact SETI, METI, and other approaches to finding intelligent life considered to be allied fields, SETI is better understood as an interdisciplinary field with many subfields, approaches, and components.[5,6,12]

In particular, Robert Bradbury [5] has argued for a broad view of SETI between two extremes: "orthodox SETI," or radio communication SETI, and a "Dysonian Approach" that searches for the extreme effects of alien life on its environment.

Here, I build on these ideas and attempt to organize the terminology and efforts of SETI within a single framework for SETI as an interdisciplinary and multi-pronged approach.

**Examples of Interdisciplinarity:**

In this broader conception of SETI, the field spans a large number of disciplines. For instance:

Just as old as radio as a branch of **radio astronomy** [11, 13] is the hunt for the effects of extraterrestrial technology on its environment [14]. Unfortunately, hopes in the 1980's that space-based **infrared astronomy** would be a quickly fruitful avenue for discovery in SETI were dashed with the discovery of the infrared cirrus background by *IRAS*, although Richard Carrigan, Jr. [8] was nonetheless able to establish some weak upper limits. Today, *WISE* and *JWST* provide new opportunities to pursue this avenue [27]. Optical and near-infrared laser SETI engages cutting-edge **optical and NIR instrumentation.**

The discovery of **exoplanets**, especially the ubiquity of apparently terrestrial bodies, has altered the astrobiological landscape across all of its subfields, and SETI is no exception [25,30]. Understanding how a civilization might interact with its terrestrial environment is a study in **Earth system science** [e.g. 16]. **Game theory** tells us that communication SETI is akin to a cooperative game where the participants cannot communicate [24], and exoplanets make excellent "focal points" in the strategy space, giving us guidance on where, when, and how to search [30].

Because SETI can involve guesses and deductions not just about alien *biology* and *evolution* (as the rest of astrobiology does) but also (alien analogs of) *psychology* and *sociology*, it is necessary to include the **social sciences** such as **anthropology** [31] to help practitioners "step out of their brains" [6] and avoid anthropocentric assumptions in their reasoning.

Understanding the spread of spacefaring life throughout the Galaxy is an exercise in galactic dynamics and stellar population synthesis, requiring participation from **galactic astrophysics**. [25] Understanding how large-scale technology feeds back on a star thermodynamically and chemically is a study in **stellar astrophysics**.

The dawn of **time-domain astronomy**, and especially the upcoming *LSST* era provide an opportunity to search for new kinds of technosignatures. The enormous quantity of data produced by upcoming time-domain missions means that even if unambiguous technosignatures are detected, our computer search algorithms may miss them because they are not looking for them. The hunt for the unexpected in enormous quantities of data requires the use of new techniques in machine learning and "big data," meaning a close relationship with **computer science**.

Since extraterrestrial technology might be enormously more advanced than humanity's, we must not neglect the possibility that evidence for it might be found in an unexpected region of the electromagnetic spectrum or even beyond it in neutrinos or gravitational waves, which puts **multi-messenger astronomy** within the domain of SETI.

Solar System SETI (the search for evidence of extraterrestrial technology in the Solar System) requires incorporating **planetary science**, **remote sensing**, and **the Earth sciences** (indeed, it is not even clear if we have sufficiently searched the *Earth* for evidence of such technology [29], and see also Schmidt & Frank 2018, in prep).

Finally, arguments about post-detection protocols, METI (the deliberate provocation of such signals via strong transmissions from Earth [17]), discovery priority, and communication of SETI results to the public require the incorporation of **media and communications**, **law** and **political science** in any mature portfolio of SETI activities.

**Jargon and Taxonomy:**

Iván Almár[1] has pointed out that difficulties in jargon arise when different forms of SETI coin and define terms narrowly, and then these terms are borrowed by practitioners in the other forms with different meanings (indeed, I have been guilty of this on many occasions). Especially if we wish to make SETI more interdisciplinary, we should ensure that our own jargon is cleaned up, and that we do not misappropriate jargon terms from the disciplines we wish to incorporate.

Indeed, the term "SETI" itself has been defined in various contexts to refer strictly to radio searches, specific NASA programs, to any search for communication, and broader searches. I concur with Almár that "**SETI**" should be the name for the entire field—after all, the field needs a name, and this is sense most consistent with the natural meaning of the term. I also endorse a slightly modified version of his definition of SETI: "the collective name of a number of activities, based on science, aimed to detect messages, signals or traces" of extraterrestrial intelligence.

Within the broader framework of science, SETI is best conceived as a **part of astrobiology**. Although the NASA Astrobiology Strategy 2015 declares that "[t]raditional SETI is not part of astrobiology," it clearly is.

The primary goal of astrobiology is the discovery of extraterrestrial life. Using Earth life as a guide, it seeks to understand how life might have arisen elsewhere, and to use this understanding detect it. One approach is the remote and *in situ* search for *biosignatures*, that is, features of creatures or their environments that distinguish them as alive.

SETI has a similar goal, except that it seeks **intelligent** life. By this, SETI practitioners use a functional definition of "intelligence" to mean "using technology whose effects we can detect." Since the term "intelligence" is loaded with anthropocentric meaning, and since explorations of the potential natures of alien intelligence is an interesting aspect of SETI, we might prefer "Search for Extraterrestrial *Technologies*" [as advocated by 15] but this jargon version of "intelligence" is so well entrenched now that I feel we should let it stand rather than attempt a rebranding.

The *reason* it is worthwhile to search for intelligent life is that it might be more detectable, because its technology can produce signatures that are both more *obvious* (a narrow-band radio signal is easier to detect than photosynthesis on part of an Earth-like planet orbiting a distant star) and more *obviously artificial* (one might always be able to "explain away" biosignatures via unlikely but plausible abiogenic means; not so for a narrow-band carrier wave).

Thus, SETI is distinguished from some other parts of astrobiology by its search for **technosignatures**. Iván Almár [2] used this term to mean any technology *other* than a communicative signal (because he was concerned with how to score the importance of various kinds of detections in SETI) but I believe the term should include any technological signature, including communication, both because that is the term's natural meaning and because the contrast with *biosignatures* helps identify SETI's place within astrobiology.

The most data-centered aspects of SETI have historically been searches for communication—primarily in the optical, near-infrared, and radio—and the most natural term for this subfield is **communication SETI**. This we can define as searches for technosignatures that involve the transmission of information through space by carriers such as photons.

The search for other effects of technology on its environment has many names. Almár[2] refers to "technomarkers," and many have referred to searches for "artifacts" (SETA), Solar System "probes", or discussed such searches in the context of "interstellar archeology."[7,9] Milan Ćirković [10] and Robert Bradbury [5] refer to this synthesis of approaches beyond "traditional" SETI as "Dysonian SETI," in part because of Freeman Dyson's seminal paper on non-communication SETI [14].

Rather than define such searches by what they are *not* or use an eponym that refers to a small subset of them, I prefer to define the subfields by their distinguishing characteristics. So, the broad search for any sort of substance—whether megastructures [4,28] or probes or bases in the Solar System[3,18,19,23,29], or cities [20] or atmospheric pollution [22] on other planets—would be **artifact SETI** (although the distinction between this and communication SETI might not always be clean, since structures might communicate information [4] and communication might take place via a physical artifact.) Under this umbrella we would then have **waste-heat SETI**, **probe SETI**, **Solar System SETI**, and so on as appropriate.

The deliberate attempt to elicit a response by *sending* a message has been called "**active SETI**" and "messaging to extraterrestrial intelligence" (**METI**). However controversial [17] this may be, this strategy is properly considered part of SETI in my taxonomy, and I think both terms are appropriate names for it.

Recently, some have begun searching for astrophysical exotica as part of SETI, along the lines of Davies' [12] suggestion that alien life might be exhibited as "**nature plus**." Examples include disappearing stars [33], unphysical stellar pulsational modes [21], or anomalous binary stars [32]. I like the term "nature plus", and find the more obvious "natural SETI" to be too vague, but I feel that the name for this field should arise organically from its practitioners (hopefully in the near future). Regardless, I would classify this as part of the search for technosignatures.

I illustrate the above taxonomy in Figure 1.

**Other Jargon:**

The term "beacon" is a useful one and should be well defined. Iván Almár[1] favors defining a beacon as a content-free attempt at communication (a "dial tone", or simple carrier wave), however the term is usually used more broadly to mean any "We Are Here" message. The latter sense has several benefits: it is the more natural meaning of the term, it allows the term to be applied across SETI, and it directly informs search strategies.

The last point is worth elaboration: if we assume that beacons in this latter sense exist, then our SETI strategy can be focused on **Schelling points** [24,30] or aspects of mutual understanding between us and the species that wishes to be found. That is, we should define a **beacon** as a signal or artifact is *meant* to be discovered by strangers, and will therefore be both obvious and obviously artificial. This means that we should be able to use physics, mathematics, or other presumed universal concepts to guess at the forms beacons might take (a topic of many SETI papers, e.g. those proposing "magic frequencies").

Perhaps the most elusive jargon is what to call what we are looking for. Given that we have already accepted the jargon term "intelligence," it is simplest to refer to **ETIs** as the targets of the search. Other common terms are "extraterrestrial civilization" (ETC) or "advanced technological civilization" (ATC) but "**civilization**" has a jargon meaning in anthropology. If we are to incorporate anthropology into SETI then we should honor their jargon, and acknowledge that that term may unnecessarily bring along more anthropocentric meaning than necessary. After all, if we make contact with an ETI we might expect it to be composed of multiple civilizations, or even none at all. "Alien **race**" is an even more loaded term, and should be avoided.

When referring collectively to the actual creatures, I prefer the term **species** because it implies separate evolutionary descent and, by analogy with *H. sapiens* might reinforce that not all members of an ETI can be expected to share similar cultural properties (and so might help avoid the "monocultural fallacy" that so often crops up in SETI papers, [cf. 26]).

Similarly, the discussion of the **colonization** of other planets and the Galaxy connotes analogy to the colonial activities of European powers on Earth (indeed, early uses of the term in this context did so quite deliberately [31]) which is unnecessarily anthropocentric. Unless one really intends such connotations, cognates of **settle** are more appropriate.

**Summary:**

There have been many calls for a broader view of SETI than is typically projected, and the present call [6] is part of what seems to me to be a recently-achieved critical mass for this attitude.

We should consider SETI to be an interdisciplinary study that includes the humanities and social sciences, and a subfield of astrobiology that focuses on the detection of technosignatures (as opposed to biosignatures). Two major branches of SETI are communication SETI and artifact SETI (although the line between them is not always sharp), and others include METI and the search for "nature plus".

# Astrobiology

## Biosignatures

**Exoplanets:**
Atmospheric gases
Reflection spectra

**Solar System:**
Microfossils
Molecular Biomarkers

## Technosignatures (SETI)

| Communication | Artifact | "Nature-plus" | METI |
|---|---|---|---|
| NIROSETI: pulsed lasers continuous lasers | Waste heat: Dyson spheres Kardashev Type III galaxies | Disappearing stars | Strong radio trasmission |
| Radio: carrier wave broadband | Solar System: Probes Structures | "Tickling" Cepheids | Embedded messages in leaked emission |
| | Exoplanets: Pollution Waste heat | Ĝ red spirals | Voyager records Pioneer plaques |
| | Transits: Arnold megastructures | Przybylski's Star | |